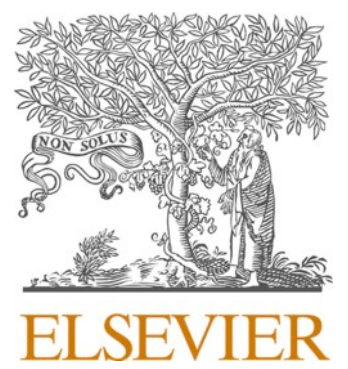

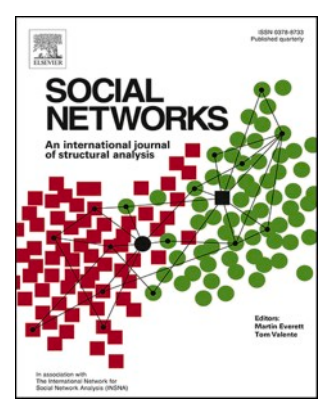

# Why distinctiveness centrality is distinctive

Andrea Fronzetti Colladon [a,b,*,1], Maurizio Naldi [a,c]

[a] Department of Civil, Computer Science and Aeronautical Technologies Engineering, Via della Vasca Navale 79, Rome 00146, Italy
[b] Department of Engineering, University of Perugia, Via G. Duranti 93, Perugia 06125, Italy
[c] Department of Law, Economics, Politics, and Modern Languages, LUMSA University, Rome 00192, Italy

ARTICLE INFO

ABSTRACT

This paper responds to a commentary by Neal (2024) regarding the Distinctiveness centrality metrics introduced by Fronzetti Colladon and Naldi (2020). Distinctiveness centrality offers a novel reinterpretation of degree centrality, particularly emphasizing the significance of direct connections to loosely connected peers within (social) networks. This response paper presents a more comprehensive analysis of the correlation between Distinctiveness and the Beta and Gamma measures. All five Distinctiveness measures are considered, as well as a more meaningful range of the α parameter and different network topologies, distinguishing between weighted and unweighted networks. Findings indicate significant variability in correlations, supporting the viability of Distinctiveness as alternative or complementary metrics within social network analysis. Moreover, the paper presents computational complexity analysis and simplified R code for practical implementation. Encouraging initial findings suggest potential applications in diverse domains, inviting further exploration and comparative analyses.

## 1. Introduction

In 2020, our publication introduced 'Distinctiveness Centrality' (DC) as a suite of 5 new measures of centrality in social networks (Fronzetti Colladon and Naldi, 2020), offering a reinterpretation of degree centrality (Freeman, 1979) tailored to pinpoint the role played by direct node connections, particularly when these occur with loosely connected neighbors. Though loosely connected, these nodes are not necessarily pertinent to the network periphery; for example, they may serve as crucial links between a network's core and periphery despite their few connections. Distinctiveness, as a metric, scrutinizes the defining characteristics of a node's direct connections.

The initial motivation and value of introducing Distinctiveness centrality stems from transferring the Term Frequency – Inverse Document Frequency (TF-IDF) concept to networks. In text mining, TF-IDF is a widely used weighting system (Ramos, 2003), for example, to determine keywords based on their frequency within a document relative to their frequency across the corpus, thus offsetting common terms (i.e., assigning a zero score to terms that appear in every document or giving low scores to terms that occur in a large number of documents). Distinctiveness, particularly in its first two formulations, represents an effort to enrich studies exploring word networks' properties and analytical methodologies. Much relevant research, indeed, converges at the intersection of text mining and social network analysis (Fronzetti Colladon et al., 2020).

This paper responds to Neal's (2024) comment, aiming to enrich the ongoing debate surrounding our metrics, shed light on their properties, and stimulate further scientific debate. We extend our gratitude to Neal for dedicating time to analyzing two of our proposed metrics and publishing a commentary on them. Additionally, we appreciate the author's mention of the Gamma (Neal, 2011) and Beta (Bonacich, 1987) centrality metrics, which bear similarities to Distinctiveness under certain circumstances.

In the following, we briefly summarize the main points of Neal's (2024) critique for the reader's convenience. Neal argued that the Distinctiveness metrics are redundant, suggesting they are minor variations of Beta and Gamma centralities (Bonacich, 1987; Neal, 2011). To demonstrate this similarity, the author conducted a correlation analysis, albeit considering only two out of the five Distinctiveness measures. Moreover, the values of the alpha parameter used in the Distinctiveness formulas were constrained and less significant than those envisioned during the parameter's conceptualization.

* Corresponding author at: Department of Civil, Computer Science and Aeronautical Technologies Engineering, Via della Vasca Navale 79, Rome 00146, Italy.
*E-mail addresses:* andrea.fronzetticolladon@uniroma3.it (A. Fronzetti Colladon), m.naldi@lumsa.it (M. Naldi).
[1] ORCID: 0000–0002-5348–9722

Given the results of this analysis, the author advised against using Distinctiveness centrality, recommending Gamma and Beta centralities instead. The justification was that these two metrics are already well-established in the literature and do not require a specialized software package for computation – as they "can be written elegantly in a matrix form that expresses the complete vector of centrality scores as a function of an adjacency matrix" (Neal, 2024, p. 7).

In Section 3, we thoroughly examine the correlations between the rankings generated by the five Distinctiveness metrics and those by Beta and Gamma centralities, surpassing the mere consideration of only the first two Distinctiveness measures – D1 and D2 (formulas provided in Section 2). In addition, we carefully explore a larger range of α parameter values within the Distinctiveness formulas. We further extend the analysis to scrutinize both weighted and unweighted networks characterized by Small-World and Scale-Free topologies. Our results show that while Distinctiveness may bear similarities to Beta and Gamma centralities in certain configurations, it yields distinct rankings. Therefore, we respectfully disagree with Neal's (2024) position advising against the use of Distinctiveness centrality. Despite their potential similarities with other measures, we maintain that our metrics operate on distinct principles and conceptualization, with Distinctiveness serving as a potential source of inspiration for scholars investigating word networks or other domains. Indeed, the significance of Distinctiveness has been illustrated in various contexts. For example, it has proven valuable for analyzing semantic networks, as also demonstrated by its integration into the Semantic Brand Score (Fronzetti Colladon, 2018) as a substitute for degree centrality (e.g., Santomauro et al., 2021; Vestrelli et al., 2024). Additionally, it was used in the mapping of technological interdependence – proving valuable in the analysis of the structure of the innovators' network (Fronzetti Colladon et al., 2025) – and the analysis of urban networks (Fronzetti Colladon et al., 2024) – providing valuable insights for the pricing strategies of gas stations –, among other applications (e.g., Silva, 2021; Yudhoatmojo, 2024).

In Section 2, we address Neal's comment regarding the elegance of code, which we believe is a less compelling argument. We also discuss the potential benefits of utilizing a software package designed to manage graph objects rather than relying on potentially unwieldy adjacency matrices. Furthermore, we demonstrate that all Distinctiveness metrics can be efficiently expressed in just a few lines of R code, similar to how the author presented Beta and Gamma centralities, using adjacency matrices as input. We also explore the asymptotic complexity of each metric, emphasizing that this aspect is more pertinent to their practical application than the current structure of the R code.

Lastly, we agree with Neal (2024) on the perspective that Distinctiveness measures, contingent upon the selection of their α parameter, can be construed as metrics of power[2] – potentially reflecting a social actor's ability to exert influence or control over, or exploit, poorly-connected neighbors, with power coming from "being connected to those who are powerless" (Bonacich, 1987, p. 1171).

This research demonstrates the absence of inherent limitations in the application of Distinctiveness centrality and encourages its continued exploration and utilization within academic research. We elaborate on our perspectives further in the remaining sections of the paper.

## 2. R code and computational complexity

Neal contends that opting for Beta or Gamma centralities over Distinctiveness centrality presents certain advantages as they "can be written elegantly in a matrix form that expresses the complete vector of centrality scores as a function of an adjacency matrix" (Neal, 2024, p. 7). However, we argue that the availability of R and Python packages tailored for Distinctiveness calculations adds significant value. These packages streamline the process by incorporating essential functions, such as data validation checks, and accommodate the computation of five distinct metrics rather than a single one. Moreover, they include variants of Distinctiveness metrics tailored for directed networks. In other words, in developing our Python and R packages, our primary emphasis was ensuring robustness by validating input data and delivering accurate results. We designed the packages to seamlessly integrate with graph objects from libraries like igraph (https://igraph.org) or networkx (https://networkx.org/), thus enhancing their versatility and compatibility. While there is potential for further optimization to enhance efficiency, it is important to note that any such refinements do not impede the usability or effectiveness of our metrics.

Additionally, the memory-intensive nature of handling large adjacency matrices during computation may pose resource constraints. In contrast, utilizing graph-type objects provided by libraries such as igraph or networkx could offer more efficient storage and operations for large-scale graphs. Therefore, there are potential efficiency gains from leveraging these well-known and widely-used packages that should be considered and explored further.[3]

In Table 1, we show the R code proposed by Neal (2024) to calculate Beta and Gamma centralities that we consider for our analysis, together with the mathematical formulas for these two metrics, to support the reader, although their detailed discussion is deferred to previous papers (Bonacich, 1987; Neal, 2011, 2024).

In the above formulas, $A$ is an adjacency matrix, 1 is a column vector of 1 s, and $\gamma$ is the tuning parameter of Gamma centrality, which is used to assign higher scores to nodes that are connected to well-connected neighbors (if $\gamma > 0$) or to nodes connected to poorly-connected neighbors (if $\gamma < 0$). $I$ is an identity matrix, *inv* is the matrix inverse function, and $\beta$ is the tuning parameter of Beta centrality, which controls whether connections to well-connected or poorly-connected nodes lead to higher Beta centrality scores, similar to the tuning parameter of Gamma centrality.

Should we choose to reimagine the code for computing all the five Distinctiveness centrality metrics in R, we could achieve this with the following functions, which maintain an "elegant" structure akin to the one proposed for Beta and Gamma centralities by Neal (2024). We are presenting this additional R code solely to address the author's critique of our use of a software package. However, the following code (as well as the code for Gamma centrality) may still encounter problems, e.g., when dealing with networks that include isolates. Therefore, we keep recommending using the distinctiveness package, which offers more robust code to prevent errors and handle such cases effectively. Table 2 presents the five Distinctiveness centrality formulas for undirected networks (Fronzetti Colladon and Naldi, 2020) and a new R code for their computation that takes an adjacency matrix as input, considering networks without isolates (for which Distinctiveness would be zero). The proposed functions essentially serve as an analog to utilizing the distinctiveness package in R (https://github.com/iandreafc/distinctiveness-R), skipping normalization. However, a notable distinction lies in the input structure: while the package operates on *igraph* class graph objects, our suggested reformulation accepts an adjacency matrix as input. It is worth noting the equivalence of D1 and D2 measures on unweighted networks, allowing for seamless interchangeability between the two on such networks.

**Table 1**
Beta and Gamma centralities. Formulas and R code.

| Metric | Formula | | R Code |
|---|---|---|---|
| Beta | $BC = inv(I - \beta A)A1$ | (1) | B <- 0 #Set value of beta<br>I <- diag(nrow(A))<br>O <- matrix(1, nrow = nrow(A))<br>beta <- solve(I - (B * A)) %*% A %*% O |
| Gamma | $GC = A(A1)^{\gamma}$ | (2) | G <- 0 #Set value of gamma<br>O <- matrix(1, nrow = nrow(A))<br>gamma <- A %*% ((A %*% O)^G) |

[2] For the sake of simplicity, we will continue to use the term "centrality" throughout the remainder of the paper.

[3] It is worth noting that Beta Centrality is available in the igraph package through the 'power_centrality()' function. In contrast, to the best of our knowledge, Gamma Centrality has not yet been implemented in any package.

**Table 2**
Distinctiveness Centrality. Formulas and R code.

| Metric | Formula | | R Code |
|---|---|---|---|
| D1 | $D1(i) = \sum_{\substack{j=1 \\ j\neq i}}^{n} w_{ij}\log_{10}\frac{n-1}{g_j^{\alpha}}$ | (3) | `alpha <- 1 #Set value of alpha`<br>`N <- nrow(adj_matrix)`<br>`degrees <- colSums(adj_matrix != 0)`<br>`d1 <- adj_matrix %*% log10((N - 1) / degrees^alpha)` |
| D2 | $D2(i) = \sum_{\substack{j=1 \\ j\neq i}}^{n} \log_{10}\frac{n-1}{g_j^{\alpha}} I_{(w_{ij}>0)}$ | (4) | `alpha <- 1 #Set value of alpha`<br>`N <- nrow(adj_matrix)`<br>`degrees <- colSums(adj_matrix != 0)`<br>`d2 <- (adj_matrix != 0) %*% log10((N - 1) / degrees^alpha)` |
| D3 | $D3(i) = \sum_{\substack{j=1 \\ j\neq i}}^{n} w_{ij}\log_{10}\frac{\sum_{\substack{k,l=1 \\ k\neq l}}^{n}\frac{w_{kl}}{2}}{\left(\sum_{\substack{k=1 \\ k\neq j}}^{n} w_{jk}^{\alpha}\right) - w_{ij}^{\alpha} + 1}$ | (5) | `alpha <- 1 #Set value of alpha`<br>`numerator <- sum(adjacency_matrix[upper.tri(adjacency_matrix)])`<br>`denominator <- rowSums(adjacency_matrix^alpha) - adjacency_matrix^alpha + 1`<br>`d3 <- colSums(adjacency_matrix * log10(numerator / denominator))` |
| D4 | $D4(i) = \sum_{\substack{j=1 \\ j\neq i}}^{n} w_{ij}\frac{w_{ij}^{\alpha}}{\sum_{\substack{k=1 \\ k\neq j}}^{n} w_{jk}^{\alpha}}$ | (6) | `alpha <- 1 #Set value of alpha`<br>`numerator <- adjacency_matrix * (adjacency_matrix^alpha)`<br>`denominator <- rowSums(adjacency_matrix^alpha)`<br>`d4 <- colSums(numerator / denominator)` |
| D5 | $D5(i) = \sum_{\substack{j=1 \\ j\neq i}}^{n} \frac{1}{g_j^{\alpha}} I_{(w_{ij}>0)}$ | (7) | `alpha <- 1 #Set value of alpha`<br>`degrees <- colSums(adj_matrix != 0)`<br>`d5 <- (adj_matrix != 0) %*% (1 / degrees^alpha)` |

In the above formulas, $n$ represents the number of network nodes, $g_j$ denotes the degree of node $j$, $w_{ij}$ signifies the weight of the edge connecting nodes $i$ and $j$, and $I(w_{ij} > 0)$ stands as a function that equals 1 if the weight of the edge linking $i$ and $j$ exceeds zero, and 0 otherwise (this function indicates the presence or absence of an edge between $i$ and $j$, as we only consider positively weighted networks). Lastly, $\alpha$ serves as the tuning parameter employed for the Distinctiveness centrality metrics, with higher values of $\alpha$ indicating a higher penalization of connections toward well-connected neighbors.

It is worth noting that D5 and Gamma centralities produce the same scores when the adjacency matrix is binary (i.e., the network is unweighted) and $\alpha = -\gamma$ (see our discussion in Section 3).

### 2.1. Computational complexity

So far, we have reformulated the R code to compute the five Distinctiveness metrics. However, as previously mentioned, we believe that computational time and efficiency in the use of computer memory are the most critical factors to consider when conducting an analysis – for example, avoiding large adjacency matrices and using more efficient graph objects. The elegance or brevity of the code is less important, especially when software packages that allow for one-line computation of these metrics are available. Accordingly, we provide an analysis of the asymptotic complexity of all the metrics discussed so far, including Distinctiveness, Beta, and Gamma centrality.[4]

It is to be noted that each metric requires the computation of the node degree. This is explicitly considered in all the Distinctiveness metrics, while it is embodied in the $A1$ term in the Beta and Gamma centralities. When the degrees for all nodes are computed beforehand, they can be subsequently retrieved from a look-up table. The computational complexity of node degree computation is $O(n^2)$, as it involves summing all the $n$ terms in each row of the weight matrix for all the $n$ rows.[5] Since this task has to be carried out for all the centrality metrics we are considering, we can view it as a lower bound for the computational complexity of these metrics. In the following, we assume that the computation of the degree has been carried out beforehand and will evaluate the computational complexity of the other operations. If that subsequent complexity should result lower than $O(n^2)$, the overall complexity will anyway be $O(n^2)$.

First, we introduce some notations that we will use for all the metrics. We indicate the number of nodes in the network as $n$, while $k$ is the number of digits of the numbers for all the quantities involved (for simplicity, we do not make differences in the numerical resolution among, e.g., node degrees, $\alpha$ values, weights, etc.). Several metrics require the computation of a logarithm function. For this elementary function, we assume that the computational complexity is $O(M(k)\log k)$ following Brent (1976), where the Arithmetic-Geometric mean method is employed (Brent and Zimmermann, 2010), and $M(k)$ is the computational effort for multiplication. The same can be said for exponentiation. As to the multiplication computational effort, it depends on the algorithm to be used, going from $O(k\log k)$ with the Harvey-Hoeven algorithm (Harvey and Van Der Hoeven, 2021) to $O(n^{1.585})$ with Karatsuba's algorithm. Please refer to Bernstein (2001) for a survey of multiplication algorithms. For the time being, since multiplications occur in all metrics, we will not specify their computational effort. In the following, we consider the computational complexity of computing each metric for all the nodes in the network, and also that $n > k$.

***D1.*** We have to raise the mode degrees of all nodes to the power of $\alpha$, carry out $n$ divisions (for which we can assume the same computational effort as for multiplications), $n$ logarithm computations, $n^2$ products by the $w_{ij}$ coefficients and $n$ sums. The overall complexity is then

$$
\begin{aligned}
C(D1) &= nO(M(k)\log k) + nO(M(k)) + nO(M(k)\log k) + n^2O(M(k)) + nO(k) \\
&= n^2O(M(k))
\end{aligned}
\tag{8}
$$

***D2***. The difference with respect to D1 is that we do not have to multiply each logarithmic term by the arc weights. The computational complexity is then

[4] As already discussed in past research, the computational complexity of Gamma centrality is lower than that of Beta centrality (Neal, 2013).

[5] The complexity of degree computation arises from using the adjacency matrix as the sole representation of the graph. If an edge list were also available, the computation would require $O(m)$ operations.

$$C(D2) = nO(M(k)\log k) + nO(M(k)) + nO(M(k)\log k) + nO(k) = nO(M(k)\log k) \tag{9}$$

However, we must consider the computation of node degrees so that the overall complexity, as far as the dependence on the network size is concerned, is again $O(n^2)$.

***D3.*** Here, we have to raise the weights $w_{ij}$ to the power of $\alpha$ and also sum them. The computational effort is

$$C(D3) = n^2O(M(k)\log k) + nO(k) + nO(k) + nO(k) + n(M(k)) + n^2O(M(k)\log k) + n^2O(M(k)) + nO(k) = n^2O(M(k)\log k) \tag{10}$$

***D4.*** The major effort here is to raise the weights $w_{ij}$ to the power of $\alpha$ and also sum them. The computational effort is

$$C(D4) = n^2O(M(k)\log k) + nO(k) + n^2O(M(k)) + n^2O(M(k)) + nO(k) = n^2O(M(k)\log k) \tag{11}$$

***D5.*** The major effort here is to raise the node degrees to the power of $\alpha$, compute their reciprocals, and sum them. The computational effort is

$$C(D5) = nO(M(k)\log k) + nO(M(k)) + nO(k) = nO(M(k)\log k) \tag{12}$$

Again, we must consider the lower bound represented by the computation of the node degrees. As to the dependence on the network size, the computational complexity is then $O(n^2)$.

### 2.1.1. Beta centrality

Assuming that the product $A1$, leading to the node degrees, has already been carried out, this metric involves the inversion of a matrix, which is probably the most relevant operation, and a multiplication. If we consider the inversion to be carried out through Gauss-Jordan elimination, its computational complexity is $O(n^3)$. Other algorithms achieve a slightly lower exponent of $n$. The computational effort is

$$C(BC) = n^2O(M(k)) + nO(k) + nO(M(k)) + n^2O(k) + O(n^3) = O(n^3) \tag{13}$$

### 2.1.2. Gamma centrality

This metric involves again the computation of $A1$ and the exponentiation of the resulting vector plus a matrix-by-vector multiplication. The computational effort is

$$C(GC) = n^2O(M(k)) + nO(k) + nO(M(k)\log k) + n^2O(M(k)) + nO(k) = n^2O(M(k)) \tag{14}$$

A brief comparison of the metrics shows that Beta centrality is the computationally most expensive one, growing with $n^3$. A lower cost is required for metrics D1, D3, D4, and Gamma, whose cost grows with the square of the number of nodes. The computational complexity of metrics D2 and D5 would grow linearly with the number of nodes if the node degrees were available beforehand. If that is not the case, their computational complexity grows again as $O(n^2)$.

## 3. Beta, gamma, and distinctiveness: a new comparison

Neal (2024) examined only two of the five Distinctiveness centrality metrics, providing a limited perspective compared to a comprehensive analysis of the entire set. Specifically, their paper outlines the formula for D1, which aligns with D2 in unweighted networks, and compares them with Beta and Gamma centralities. In the ensuing discussion, we present a thorough examination encompassing all five metrics, distinguishing between weighted and unweighted networks. Notably, we employ D2, D3, and D5 for unweighted networks, as they are tailored for such contexts, while using D1, D3, and D4 for weighted networks.

Furthermore, our analysis includes correlation values between Gamma and Beta centralities, which were not elucidated in Neal's paper.

To ensure consistency across different metrics, we adhere to formulas 4 and 5 presented in Neal's (2024) paper, which they suggested to harmonize the parameters of the different metrics. However, since our emphasis lies on Distinctiveness centrality, we illustrate the variations in correlations as the parameter $\alpha$ changes. Consequently, we establish the parameter $\gamma$ for Gamma centrality as $\gamma = -\alpha$, while defining the parameter $\beta$ for Beta centrality as $\beta = \left(\frac{2}{e^{\alpha}} - 1\right) \times \frac{1}{\lambda_1}$, where $\lambda_1$ represents the largest eigenvalue of the adjacency matrix.

While Neal (2024) only provides a short explanation of these relationships, we notice that they may derive from the metric D5 when we have a non-weighted network. In fact, the product $A1$ appearing in the Gamma centrality is the vector of degree values

$$A1 = \begin{bmatrix} w_{11} & \cdots & w_{1n} \\ \vdots & \ddots & \vdots \\ w_{n1} & \cdots & w_{nn} \end{bmatrix} \begin{bmatrix} 1 \\ \vdots \\ 1 \end{bmatrix} = \begin{bmatrix} \sum_{j=1}^{n} w_{1j} \\ \vdots \\ \sum_{j=1}^{n} w_{nj} \end{bmatrix} = \begin{bmatrix} g_1 \\ \vdots \\ g_n \end{bmatrix} \tag{15}$$

so that the Gamma centrality can be written as

$$GC = A \quad (A1)^{\gamma} = \begin{bmatrix} w_{11} & \cdots & w_{1n} \\ \vdots & \ddots & \vdots \\ w_{n1} & \cdots & w_{nn} \end{bmatrix} \begin{bmatrix} g_1^{\gamma} \\ \vdots \\ g_n^{\gamma} \end{bmatrix} = \begin{bmatrix} \sum_{j=1}^{n} w_{1j} g_j^{\gamma} \\ \vdots \\ \sum_{j=1}^{n} w_{nj} g_j^{\gamma} \end{bmatrix} \tag{16}$$

which is exactly the D5 metric when $w_{ij} \in \{0, 1\}$ so that $w_{ij} = I(w_{ij} > 0)$ and $\gamma = -\alpha$.

However, this relationship is valid for D5 only. As can be seen from the formulas for the other metrics, a similar equivalence does not apply to D1 through D4. Hence, any conclusion based on the proposed harmonization formulas to make the metrics comparable should be considered with caution. Having said that, in order to allow for a direct comparison with the analyses carried out by Neal (2024), we adopt those formulas throughout this paper.

In formulating Distinctiveness centrality (Fronzetti Colladon and Naldi, 2020), the parameter $\alpha$ was conceived to offer flexibility, allowing it to deviate from its standard value of 1. This deviation enables the penalization of connections to highly connected nodes to a greater extent. Thus, we suggested that $\alpha$ should be greater than or equal to one.

We highlighted the case $\alpha > 1$ because the contributions of the logarithmic term to D1, for example, may be negative in that case, while they are all positive when $0 < \alpha < 1$. What happens when the contributions are all positive is, however, that the nodes with a higher degree contribute less to the final score than the nodes with a lower degree. If we consider two nodes $k$ and $m$, with degrees $g_k > g_m$, we see that their contributions to D1 are such that $\log\frac{n-1}{g_k^{\alpha}} < \log\frac{n-1}{g_m^{\alpha}}$, regardless of the value of $\alpha > 0$. Hence, the general principle that nodes with a higher degree are penalized in that metric is maintained even if $0 < \alpha < 1$.

In addition, we can notice that the first derivative of D1 is always negative and does not depend on $\alpha$, so there is no discontinuity when $\alpha$ crosses the border value 1:

$$\frac{\partial \quad D1}{\partial \alpha} = -\frac{1}{\ln 10} \sum_{\substack{j=1 \\ j \neq i}}^{n} w_{ij} \ln g_j < 0 \tag{17}$$

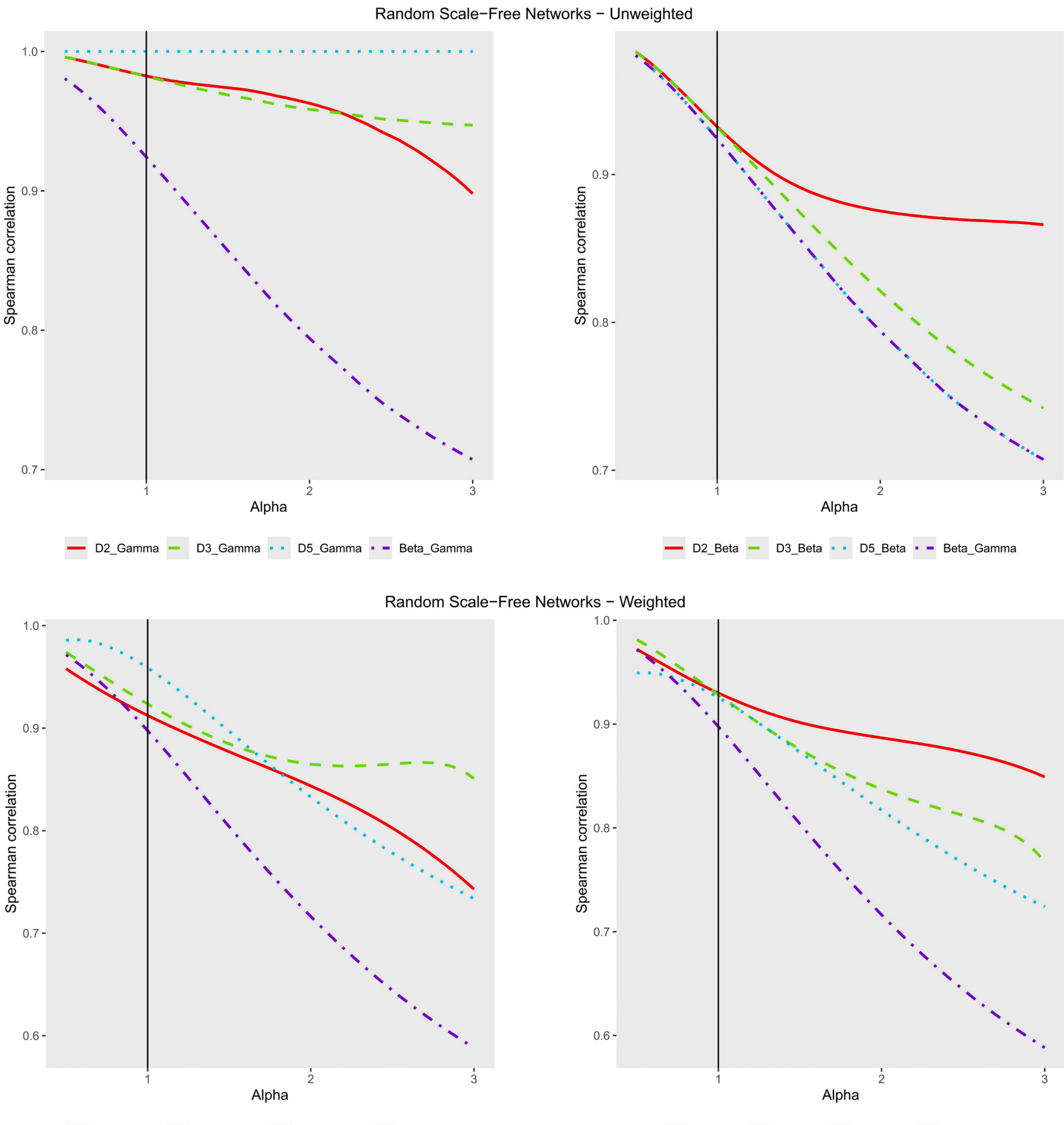

**Fig. 1.** Spearman's correlations – Scale-Free Networks.

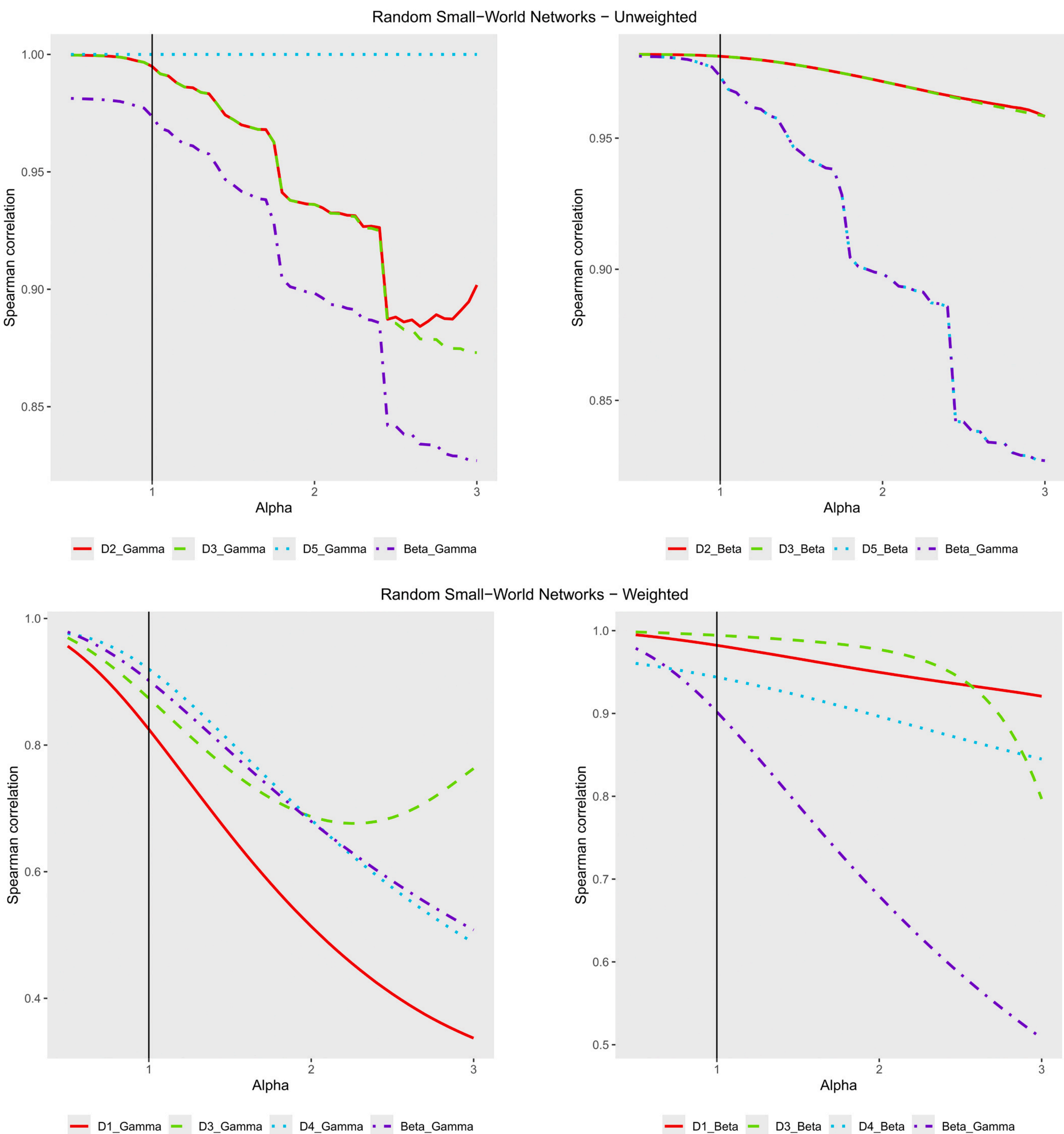

**Fig. 2.** Spearman's correlations – Small-World Networks.

Neal (2024) also wonders why we should choose larger values of $\alpha$. Well, larger values of $\alpha$, being used in the exponent of the degree of node $j$ in the formula for D1, amplify the differences in the contributions brought by high-degree and low-degree nodes, respectively. However, as shown by formula 17, D1 overall value decreases linearly when $\alpha$ grows.

Consequently, we revised the range of variation for $\alpha$ when comparing the metrics, selecting it from 0.5 to 3 – though higher values could have been considered, we constrain it to 3 for conciseness. Unlike Neal (2024), we do not normalize the metrics as it is irrelevant to the computation of correlations.

Figs. 1 and 2 illustrate the outcomes of our analysis, carried out using Spearman's correlation. In all our analyses, the correlation scores would be lower if Kendall's $\tau$ were used instead of Spearman's $\rho$; e.g., for allowing comparison with the results presented by Schoch et al. (2017). Kendall's correlation scores can be easily obtained from the code we have shared alongside this paper.

We computed the average correlations among the different metrics on randomly generated Small-World and Scale-Free networks using the igraph package in R. For each configuration (Scale-Free vs. Small-World and Weighted vs. Unweighted), we generated 200 random networks each containing 1000 nodes.[6] The code required to replicate these findings and the other analyses presented in this paper is available at https://github.com/iandreafc/distinctiveness_comparisons.

In the figures, we have arranged separate plots for the correlations of Distinctiveness with Gamma and Beta centralities to enhance readability. Fig. 1 illustrates the average Spearman's correlations observed for Scale-Free networks, while Fig. 2 focuses on Small-World networks. Across all scenarios, we notice variability in the correlations among the different metrics, giving evidence to each measure's capacity to assign distinct importance scores to network nodes.

The vertical line denoting an $\alpha$ value of 1 serves as a reference point for meaningful consideration of the parameter. As $\alpha$ increases, we observe a decline in correlation values, highlighting an increasing divergence between metrics and the adaptability inherent in Distinctiveness centrality to accommodate varying interpretations with changes in the $\alpha$ parameter. As expected, in unweighted networks, we observe a maximal correlation between Gamma and D5 – which produce the same scores for $\gamma = -\alpha$. Additionally, correlations between D2 and Beta, D2 and Gamma and D3 and Gamma, and sometimes D3 and Beta, remain relatively high. However, these values may decrease when extending the analysis to higher alpha values.[7] Importantly, this is not always the case, and results can vary depending on network topology. Our aim is not to generalize these results but rather to demonstrate the variability in correlation values across metrics based on the chosen alpha parameter, network topologies, and arc weights – see the Appendix for another illustrative analysis conducted on random graphs generated using the Erdos-Renyi model (Erdos and Renyi, 1959). Therefore, it is challenging to designate one metric as the substitute for another. Each metric could showcase an optimal use case, and we advocate for future research in this direction. An exception lies in the equal scores produced by D5 and Gamma centrality in unweighted networks. Conversely, other Distinctiveness measures diverge from Gamma and Beta centralities, particularly with increasing alpha values. The ability to easily adjust the alpha parameter, coupled with the fact that it encompasses a set of five metrics, endows Distinctiveness with unique flexibility, rendering it potentially valuable for diverse applications.

Lastly, the figures reveal a degree of correlation between Gamma and Beta centralities, which is not unexpectedly high in many cases (see also the Appendix). In general, significant and high correlations can be observed between many measures of network centrality (or power), such as degree, betweenness, eigenvector, and closeness centrality (Schoch et al., 2017). However, this does not support choosing one or two metrics and disregarding the others. While often correlated, these diverse metrics succeed in capturing distinct facets of social actors' positions. Moreover, they may demonstrate varying degrees of explanatory power concerning external variables (i.e., centrality effects), such as employee performance levels within a company (Wen et al., 2020).

## 4. Discussion and conclusion

In this paper, we presented a more comprehensive examination of Distinctiveness centrality in comparison with Beta and Gamma centralities than the one offered by Neal (2024). This involved exploring more appropriate values for the $\alpha$ parameter, incorporating all five Distinctiveness metrics, distinguishing between weighted and unweighted networks, and including the correlation between Beta centrality and Gamma centrality.

Our findings indicate that, except for the correlation between D5 and Gamma centrality in unweighted networks, the remaining correlations demonstrate significant variability and tend to decrease as the alpha parameter increases. In particular, the lowest correlations (as alpha increases), between Distinctiveness and Gamma and Beta centralities, vary by network topology, and our experiments show the following results. In unweighted scale-free networks, D3 and Beta, as well as D5 and Beta, exhibit the lowest correlations, while in weighted scale-free networks, it is D4 and Beta, D4 and Gamma, and D1 and Gamma. D5 and Beta have the lowest correlations for unweighted small-world networks, while in weighted versions, it is D1 and Gamma, along with D4 and Gamma. In unweighted Erdos-Renyi graphs (see the Appendix), the lowest correlations are observed between D2 and Gamma, and D2 and Beta. In their weighted counterparts, these are D1 and Gamma, D1 and Beta, D3 and Gamma, and D3 and Beta. This variability, which could be further investigated and justified in future research, demonstrates that Distinctiveness can produce scores that differ from those generated by Beta and Gamma centralities, positioning it as an alternative or complementary set of metrics. This addresses Neal's (2024) first critique, in which they questioned the novelty of our metrics.

Additionally, in Section 2, we presented the formulas accompanied by simplified R code and an analysis of the computational complexity of all the metrics discussed in this paper. In particular, we assessed their asymptotic computational complexities and found that Beta centrality exhibits the highest complexity at $O(n^3)$. The complexity of Gamma centrality grows with the square of the number of nodes, similar to all Distinctiveness centrality metrics when node degrees are not precomputed. If node degrees are available beforehand, the computational complexity of metrics D2 and D5 instead grows linearly with the number of nodes.

All these findings support the viability of Distinctiveness as either an alternative or complementary set of metrics to Beta, Gamma, and other conventional measures of centrality (or power) within social networks. Distinctiveness has an additional advantage in that it comprises a set of five metrics, all based on a shared conceptualization. This is significant per se, as it connects to the study of words and networks, setting it apart from the conceptualizations of Beta and Gamma centralities. Furthermore, having five metrics increases application flexibility beyond what a single metric can offer.

There is also a potential limitation worth considering when using

[6] We used the R igraph package for generating the random networks with the following functions sample_pa(n = 1000, $m$ = 2, directed = FALSE) and sample_smallworld(dim = 1, size = 1000, nei = 2, $p$ = 0.05). For the weighted versions, we attributed random weights in the range from 1 to 20 to each edge. In the shared code, we set random seeds to restrict our analysis to networks without isolates due to limitations in the Gamma code, which generates NaNs when isolates are present. Alternatively, one could use a function – such as this one g_no_isolates <- delete_vertices(g, V(g)[degree(g) == 0]) – to remove isolates.

[7] The code we share can be tested with various parameters of the network generation functions. Experimentation can include adjusting sizes and exploring different alpha values.

Spearman correlation. Take, for example, two hypothetical centrality measures applied to a network of three nodes, A, B, and C. Let us say Measure 1 assigns scores {A: 100, B: 99, C: 98}, while Measure 2 assigns scores {A: 100, B: 9, C: 1} on the same measurement scale. Spearman's correlation would yield a result of one, indicating perfect alignment in node rankings. However, this assessment overlooks the variability of scores. While a comprehensive discussion falls outside the scope of this paper, we incorporate a comparison of the score distributions generated by the various metrics in the Appendix and assess their distance using the Ruzicka index (Cha, 2007). It might be valuable for future research to delve more into these nuances, exploring additional methods for comparing scores derived from Distinctiveness and other centrality measures.

Similarly, future research should consider the inherent limitations of using correlation analysis to assess the redundancy of new centrality metrics. Correlations among centrality metrics, which are often high, may not necessarily reflect formal or conceptual similarities, as they can be confounded by underlying network structures (Schoch et al., 2017). Therefore, not only correlations may be questioned as the most appropriate method for assessing the redundancy of centrality metrics, but this adds to the potential concerns raised in Section 3 regarding the proposed harmonization formulas for the $\alpha$, $\beta$, and $\gamma$ parameters, which are used to compare the metrics.

Despite being a relatively recent introduction, Distinctiveness centrality metrics already demonstrate promise across various domains. For example, they have shown potential in analyzing semantic networks, urban networks, and technological interdependence relationships between sectors (Fronzetti Colladon et al., 2024, 2025; Vestrelli et al., 2024).

In terms of future research directions, it would be important to conduct comparative analyses of the explanatory capabilities of Distinctiveness metrics, Beta, Gamma, and other traditional centrality measures in diverse contexts. Such investigations could provide insights into certain observed phenomena and contribute to their understanding through empirical examination.

Certainly, the encouraging initial findings of the aforementioned studies should catalyze scholars to delve deeper into potential and possible applications of Distinctiveness rather than constraining its exploration.

Furthermore, the conceptualization of Distinctiveness holds promise as a source of inspiration for researchers interested in the study of semantic networks. By bringing the logic of TF-IDF transformation to networks, Distinctiveness centrality introduces an inherently innovative approach that has the potential to inspire researchers engaged in studies at the intersection of text mining and network analysis.

While not the primary focus of this paper, it is worth noting that Distinctiveness has also been adapted for computation on directed networks, thereby extending the logic of metrics such as in-degree and out-degree. Future research could explore these extensions and broaden the scope of comparisons discussed herein.

In conclusion, we appreciate Neal's (2024) dedication and interest in analyzing two out of five Distinctiveness metrics. In general, we believe that discussions such as the one sparked by our 2020 paper are pivotal for the advancement of science, and we are grateful for that. Nevertheless, we respectfully disagree with the authors' suggestion to forgo the utilization of Distinctiveness centrality in research. Their correlation analysis is somewhat limited, as it focuses solely on D1 and D2 and does not encompass $\alpha$ values greater than 1, which are integral to the original logic of the metrics. Moreover, the arguments regarding the elegance of Beta and Gamma centrality formulations with respect to Distinctiveness appear weak. In Section 2, we have presented a simplified R code for computing the five metrics and an analysis of their asymptotic complexity – an aspect we believe holds more relevance for researchers than considerations of programming code structure. As previously mentioned in Section 2, utilizing a package enabling metric calculation directly from an igraph object, as opposed to an adjacency matrix, offers the additional advantage of conserving computer memory resources. This approach allows for calculations even on sizable networks, facilitating analysis on standard commercial PCs – a task probably unfeasible if working with large adjacency matrices.

The exploration we undertook unveils new avenues of research regarding the potential of Distinctiveness. It also highlights how Neal's (2024) judgment may have been rendered without considering more comprehensive comparisons and broader perspectives.

Accordingly, we maintain that researchers should be afforded the freedom to conduct their own evaluations and select the metrics that best suit their research needs, whether it be Beta, Gamma, Distinctiveness centrality, or alternative metrics.

## Declaration of Generative AI and AI-assisted technologies in the writing process

While preparing this work, we used Grammarly and ChatGPT solely to refine the language. After using these tools, we reviewed and edited the content as needed. We take full responsibility for the content of the publication.

## CRediT authorship contribution statement

**Andrea Fronzetti Colladon:** Writing – review & editing, Writing – original draft, Visualization, Validation, Software, Methodology, Funding acquisition, Formal analysis, Conceptualization. **Maurizio Naldi:** Writing – review & editing, Writing – original draft, Methodology, Formal analysis, Conceptualization.

## Acknowledgments

We are very grateful to the two anonymous reviewers for their constructive feedback.

We also wish to thank Zachary Neal for sparking this scientific debate and offering valuable suggestions during the revision of this manuscript.

This work was partially supported by the University of Perugia through the program Fondo Ricerca di Ateneo 2022, Proj. "Argomentazione Astratta, Text Mining e Network Analysis per il Supporto alle Decisioni (RATIONALISTS)". The funders had no role in study design, data analysis, decision to publish, or preparation of the manuscript.

## Appendix

Fig. A1 illustrates a correlation analysis akin to the one outlined in Section 3. Here, we employed the sample_gnp(100, 0.1) function from the igraph package to generate 100 random networks, each with a size of 100 (to provide an example on smaller networks), using the Erdos-Renyi model (Erdos and Renyi, 1959). For the weighted versions, we attributed random weights in the range from 1 to 80 to each edge (to provide an example with greater variability in arc weights). As we can observe from the figure, some correlations drop fast and even become negative for $\alpha$ values bigger than two.

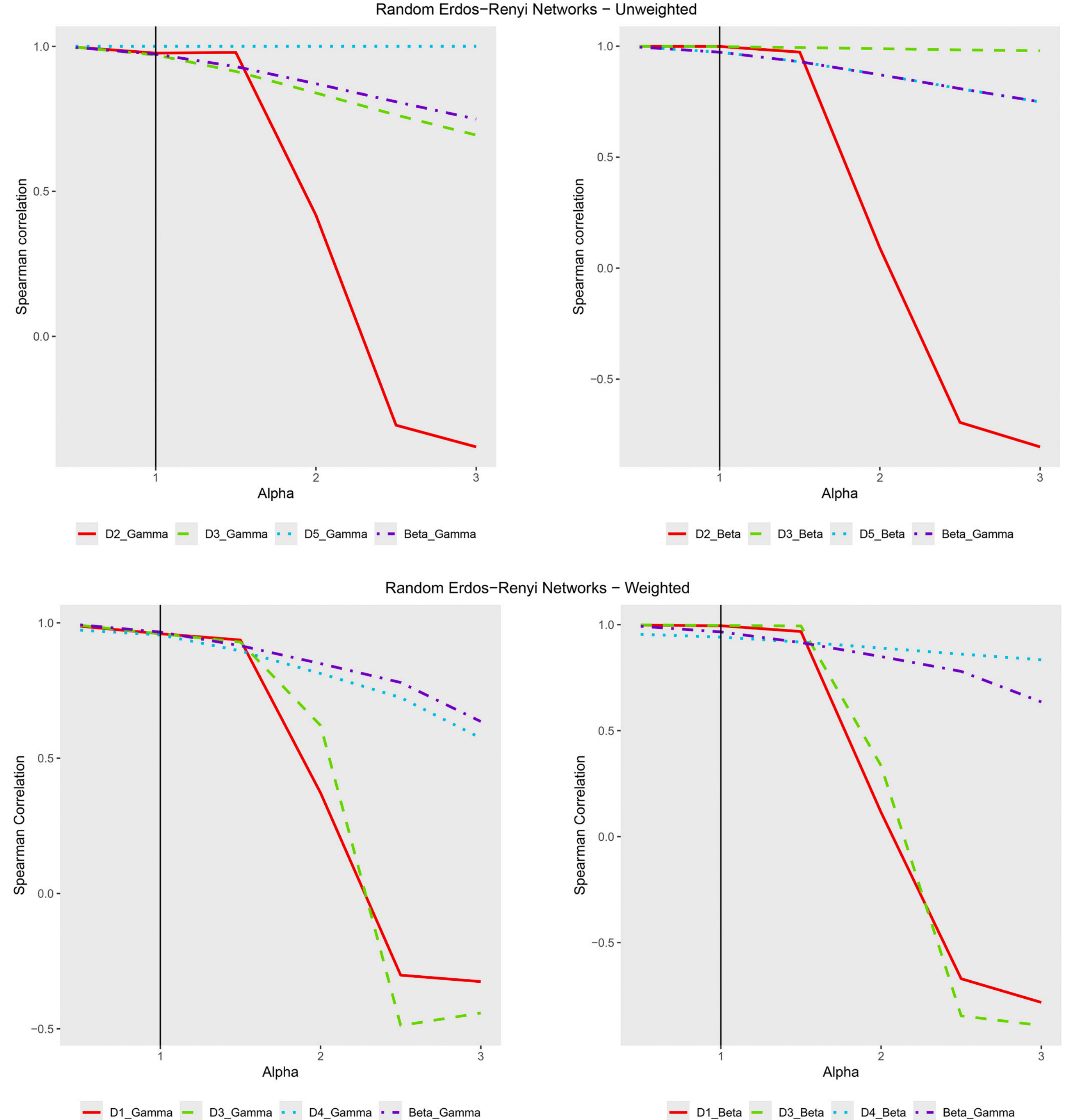

**Figure A1.** Spearman's correlations – Erdos-Renyi Networks.

In the subsequent figures, we present an analysis aimed at comparing the distributions of scores generated by the Distinctiveness metrics and Beta and Gamma centralities. This comparison extends beyond Spearman's correlations, focusing on assessing the distances between scores and their densities. As mentioned in Section 4, this analysis represents only a partial exploration and invites further investigation in future research endeavors.

We used a random Scale-Free network comprising 1000 nodes to conduct the analysis.[8] Random weights ranging from 1 to 80 were assigned to the edges to calculate the weighted version of the metrics. Each of the following figures presents density plots of the scores obtained from the metrics alongside a heatmap illustrating the values of Ruzicka's index (Cha, 2007) for comparisons between each pair of distributions.

Figs. A2, A3, and A4 show density plots and Ruzicka indices on the unweighted network for alpha parameter values of 1, 2, and 3 (with

[8] The network was generated using the R code "sample_pa(n = 1000, $m$ = 2, directed = FALSE)". The same analysis could be replicated on Small-World or Erdos-Renyi networks, by simply adapting the R code we made available.

corresponding adjustments of values for gamma and beta parameters, as discussed in Section 3). In this context, we explore Distinctiveness metrics D2, D3, and D5, tailored for unweighted networks, along with Gamma and Beta centralities.

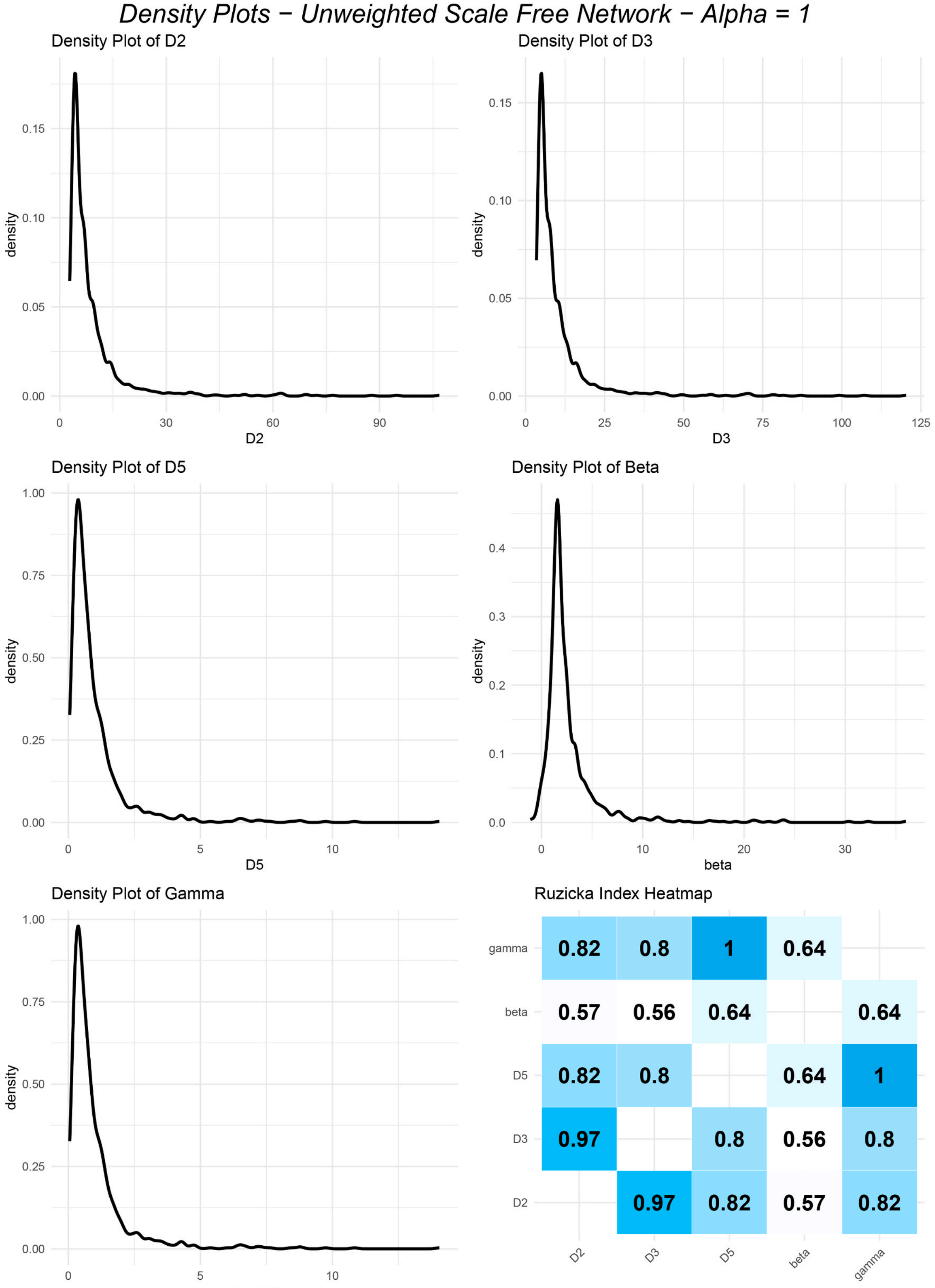

**Figure A2.** Comparing distributions. Unweighted scale-free network, alpha = 1.

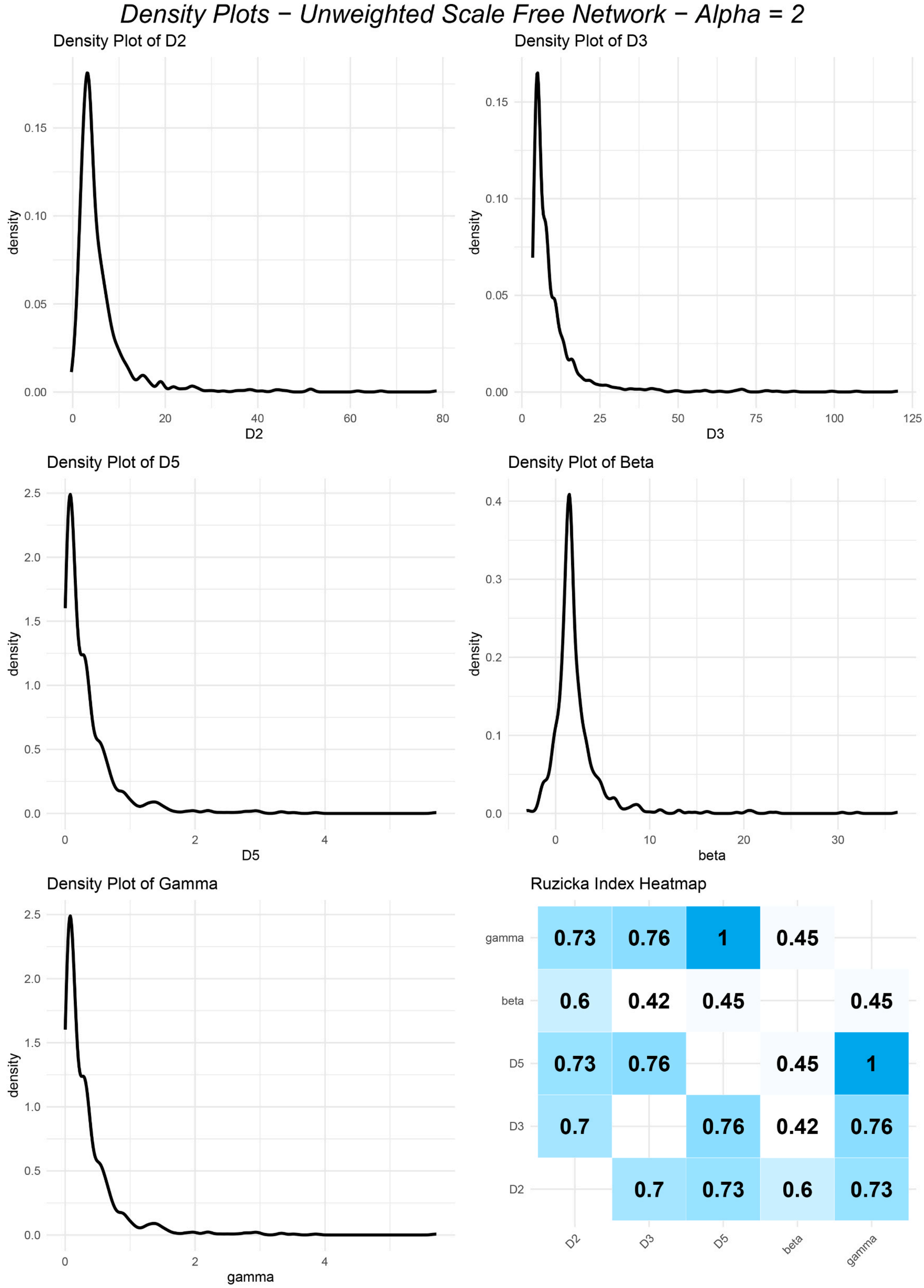

**Figure A3.** Comparing distributions. Unweighted scale-free network, alpha = 2.

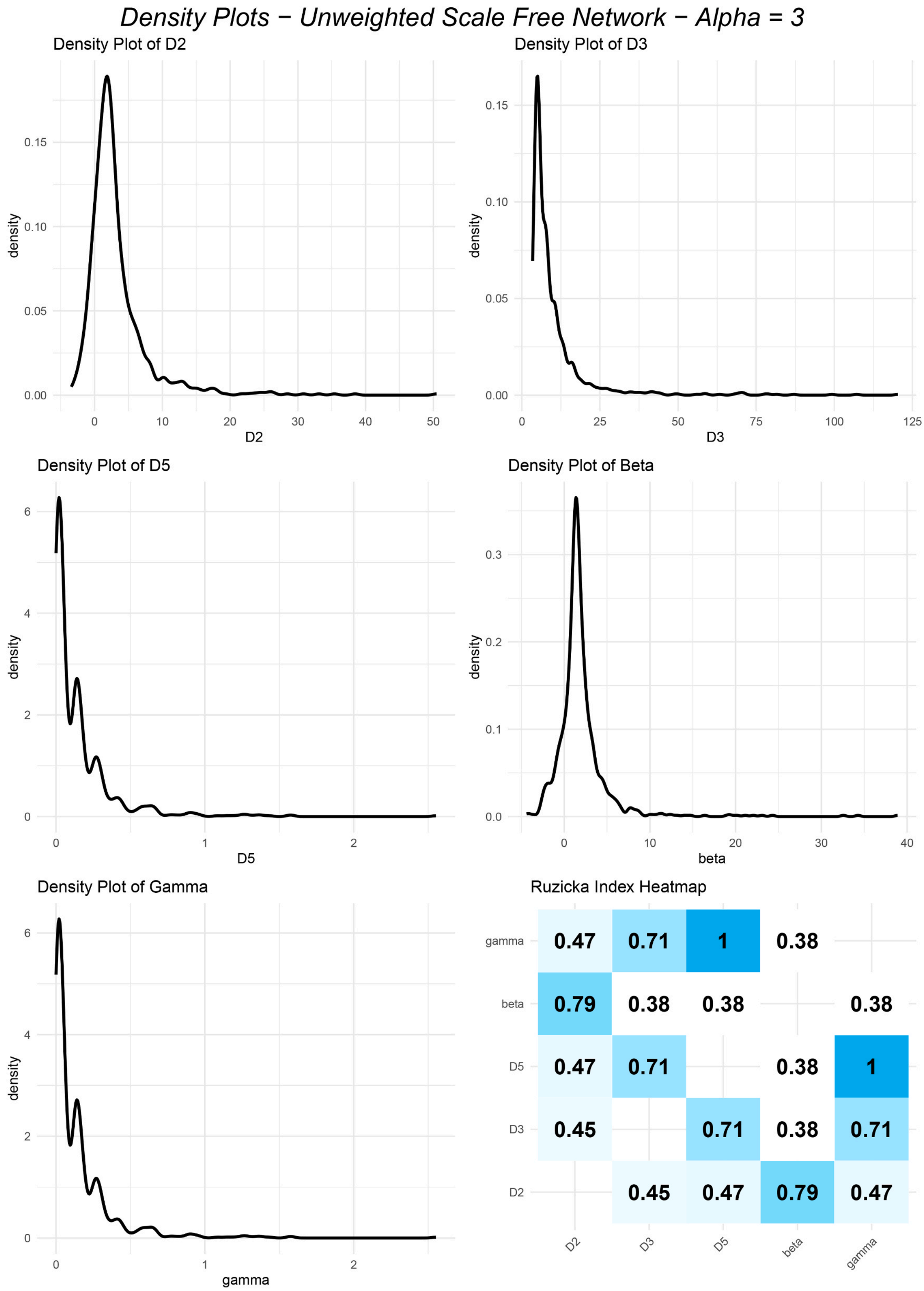

**Figure A4.** Comparing distributions. Unweighted scale-free network, alpha = 3.

Similarly, Figs. A5, A6, and A7 present the same analysis but with random weights assigned to the edges. In this scenario, we examine Distinctiveness metrics D1, D3, and D4, specifically designed for weighted networks, in addition to Gamma and Beta centralities.

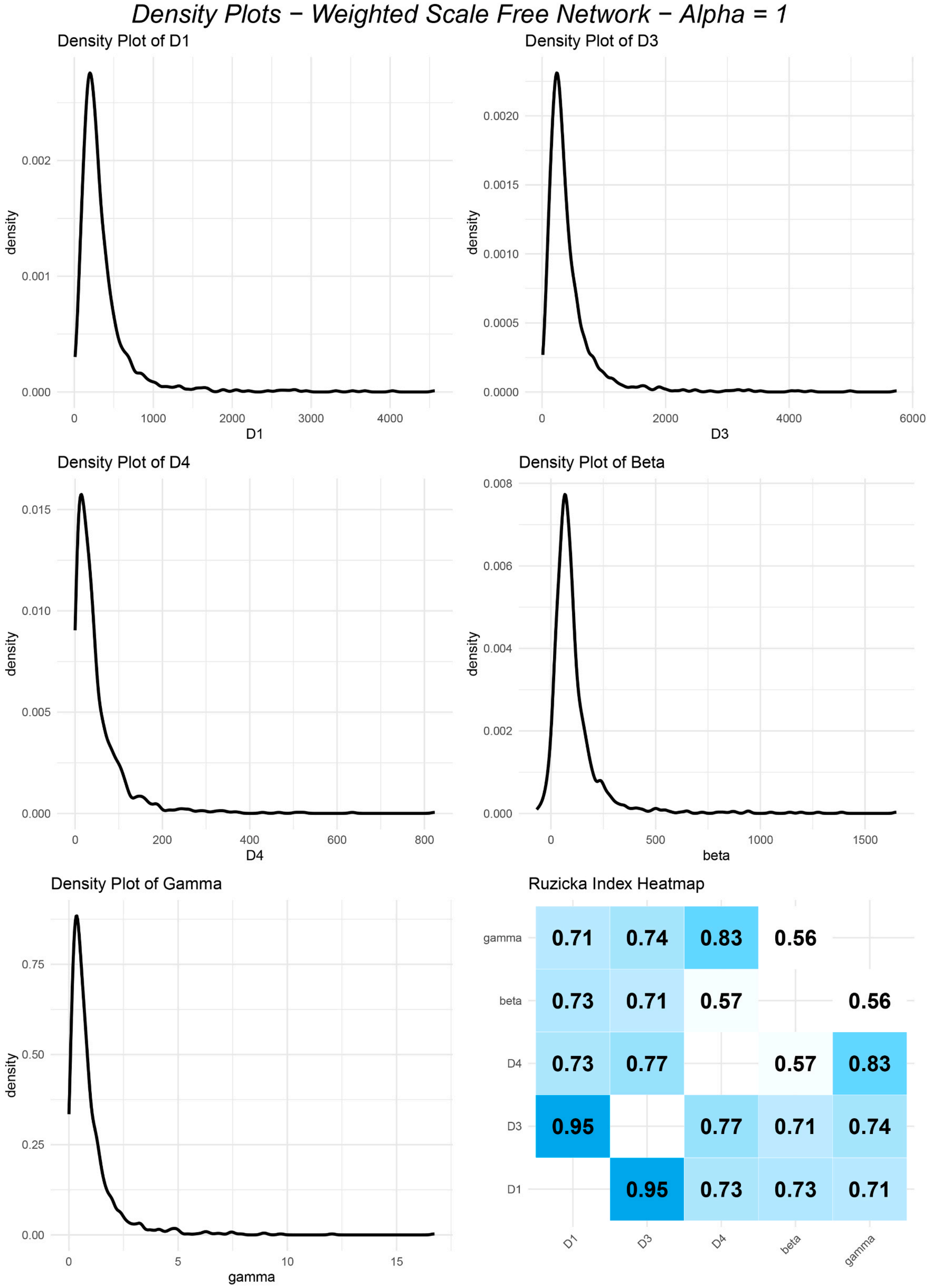

**Figure A5.** Comparing distributions. Weighted scale-free network, alpha = 1.

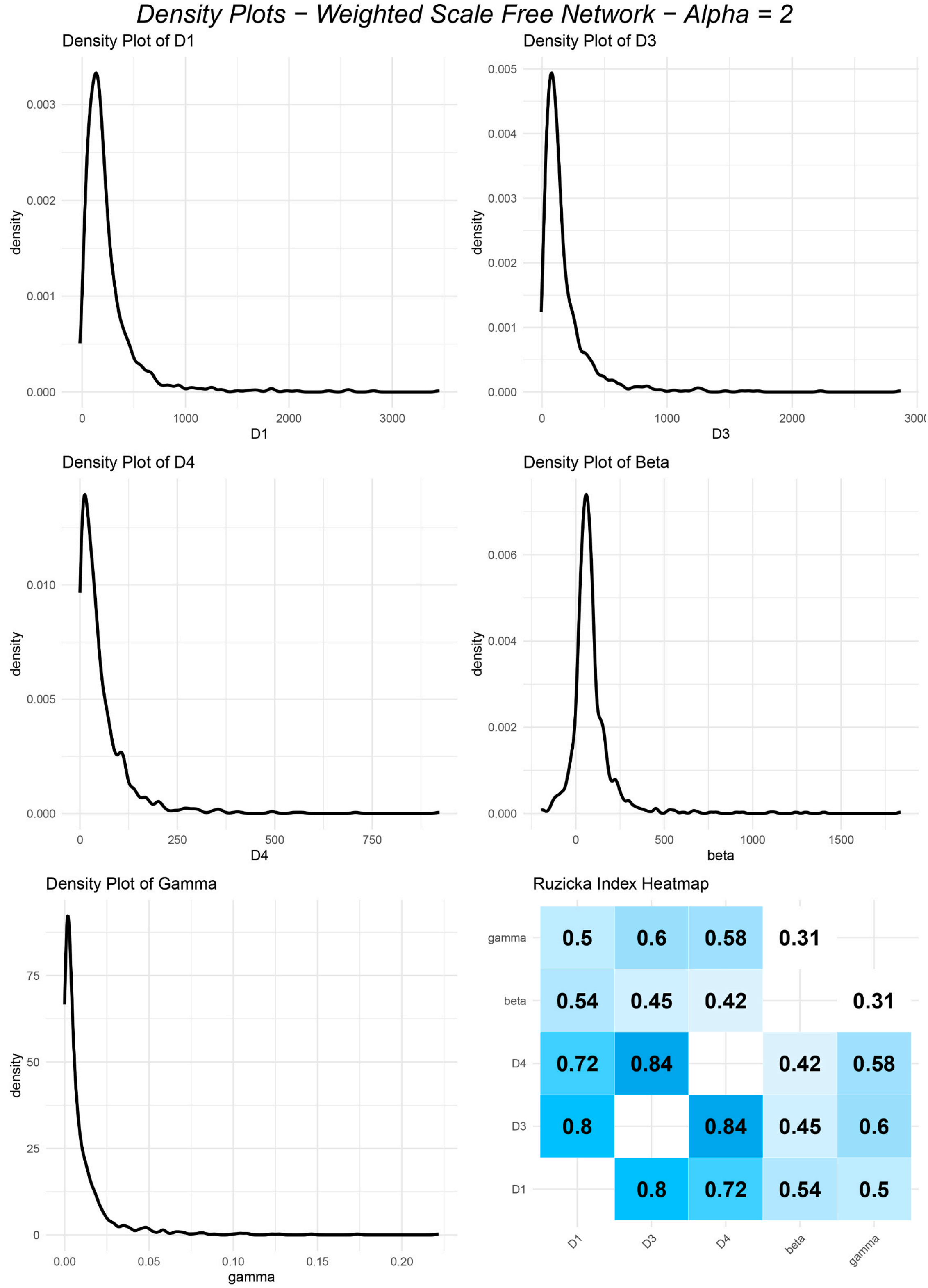

**Figure A6.** Comparing distributions. Weighted scale-free network, alpha = 2.

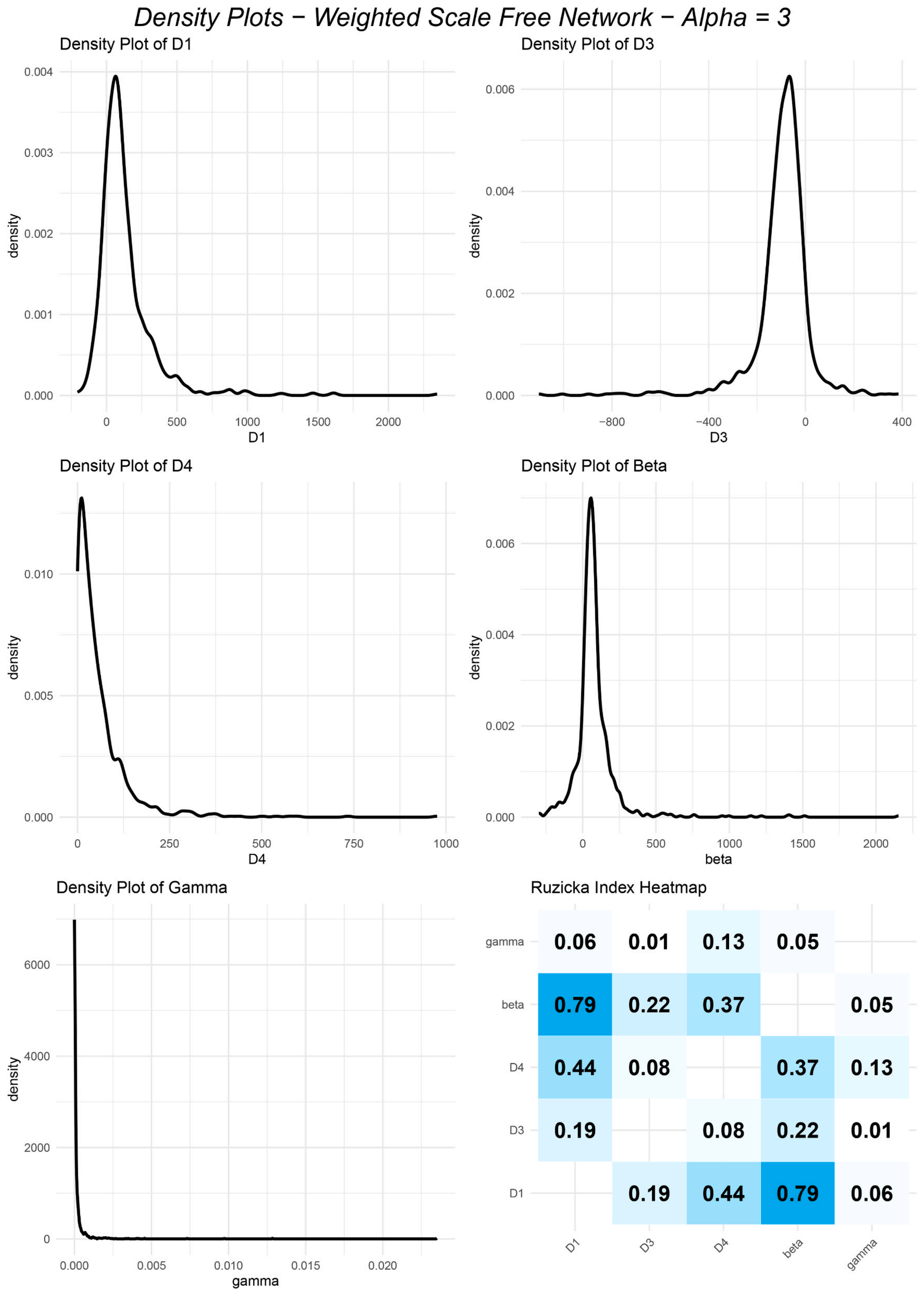

**Figure A7.** Comparing distributions. Weighted scale-free network, alpha = 3.

Our findings highlight the variability inherent in the proposed metrics, aligning with the outcomes derived from Spearman's correlations. Once again, we find perfect equivalence between D5 and Gamma centrality in unweighted networks when $\alpha = -\gamma$. Moreover, as alpha values increase, we note a decline in the average of the Ruzicka indices, indicating a widening gap among the metrics under examination. While this trend holds true for most cases, there are exceptions. For example, the distributions of Beta and D2 centrality scores on the unweighted network reduce their distance as alpha increases, as do Beta and D1 scores in the weighted network when alpha increases from 2 to 3.